# Alternative Interpretation of Lens Aberration on OTF

Yu Bai, Jiaqi Chen, Fangjie Li, Zhenming Zhao [*]

School of Science, Changchun University of Science and Technology, Jilin 130022, China

[*] Corresponding author at: School of Science, Changchun University of Science and Technology, Jilin 130022, China.

E-mail address: zhaozhenming@cust.edu.cn (Z. Zhao).

## Abstract

This paper demonstrates a method of interpreting the mechanism of aberration of optical systems based on non-Fourier transform optical transfer function (OTF). According to the parameters of object plane cosine fringe, we obtain the position on the pupil plane of the two beams of the interference pair emitted by the cosine fringes. By combining the aberrations of the lens, the phase of the cosine fringe of each interference pair in the image plane can be obtained. Then the intensity of the cosine fringe formed by all the interference pattern is superimposed to obtain the brightness distribution of the cosine fringe on the image plane. After that, the OTF of the aberration system can be obtained on the basis of the cosine fringe modulation on the image plane and the phase shift. Finally, taking the non-Fourier transform OTF theory with wave analyze the complex optical system, providing help for the optical system design and imaging quality evaluation.

## Introduction

The OTF is a method which can be used to evaluate the performance of optical imaging systems comprehensively and objectively[1, 2]. However, analyzing the imaging process in the spatial domain and combining the Fourier transform is the current method of determining an OTF[1-6]. The latest method, which is the theory of OTF without Fourier transform, aims to obtain the OTF of the system from the frequency domain imaging in an intuitive way[7,8]. It is based on the interference theory of light and the concept of the photon superposition state, which consider that the light emitted by the cosine fringe of the object plane is decomposed into a large number of interference pairs. By analyzing the utilization efficiency of the imaging system to the interference pair, we can get the modulation of the cosine fringe of the image surface, and then the OTF of the imaging system also can be obtained. We use the non-Fourier transform OTF theory to analyze the effect on OTF of aberration of the single thin lens, which shows that the non-Fourier transform OTF theory has the intuitiveness of the geometrical optics and the accuracy of wave optics. In the non-Fourier transform OTF theory, the light beam replaces the geometrical optics rays and tracks the propagation process of the light beam according to the construction of the optical system. The OTF of the imaging system is obtained in an intuitive way, which is especially suitable for analyzing the OTF of the optical imaging system composed of multi-lens. And it gives full play to the advantages of the intuitive image of the non-Fourier transform OTF theory. Since the actual optical system with aberrations, if we can extend the non-Fourier transform OTF theory to the system which has an aberration, it will not only assist in the design and analysis of optical imaging systems but also advance the OTF theory of non-Fourier transform.

## 1. Lens imaging process in frequency domain

The OTF without the Fourier transform uses the light beams instead of geometric optics. According to the concept of photon superposition states in quantum mechanics, the light wave emitted by the cosine fringe of the object plane is decomposed into a large number of interference pairs. After the interference pair is refracted through the lens, it is superimposed on the image plane to form a cosine fringe. Due to the limit of the lens aperture, some interference pairs cannot reach the image plane completely, and only some interference pairs can participate in the imaging process. This led to a drop in the modulation of the image.

Figure.1 is the process of cosine fringe imaging. In order to satisfy the paraxial condition, we only need to study the imaging process of cosine fringes near the optical axis. Assuming the spatial frequency expression of cosine fringe of the object plane is:

$$I_o(x_o, y_o) = 1 + \cos\left[2\pi(\xi_o x_o + \eta_o y_o)\right] \quad (1)$$

According to the non-Fourier transform OTF theory[8], when a large number of interference pairs are emitted by the cosine fringe of the object plane, the interference pair is refracted through the lens to the image surface and interferes to form the cosine fringe. In figure 1, $s'$ and $s''$ are two beams of an interference pair, they're going to be points $S'$ and $S''$ on the xy-plane. From the spatial frequency of the cosine fringe on the object surface $(\xi_0, \eta_0)$, we can know that the coordinate difference of points $S'$ and $S''$ on x-axis is $\lambda d_0 \xi_0$, and on y-axis the difference is $\lambda d_0 \eta_0$ (show as Fig.2). Thus, the coordinates of point $S'$ and point $S''$ can be written as $S'(x + \lambda d_0 \xi_0, y + \lambda d_0 \eta_0)$ and $S''(x, y)$. When the lens is imaging, the imaging magnification $M = \dfrac{d_i}{d_o} = \dfrac{\xi_i}{\xi_o} = \dfrac{\eta_i}{\eta_o}$, it has $d_o \xi_o = d_i \xi_i$ and $d_o \eta_o = d_i \eta_i$. Thus, $S'$ also can be written as $S'(x + \lambda d_i \xi_i, y + \lambda d_i \eta_i)$.

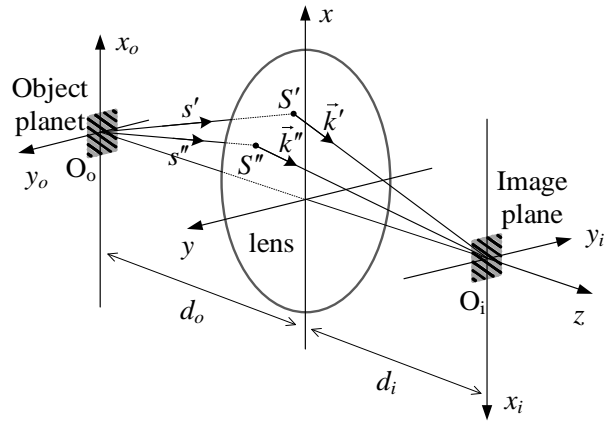

Figure 1. The interference pair propagation process during lens imaging.

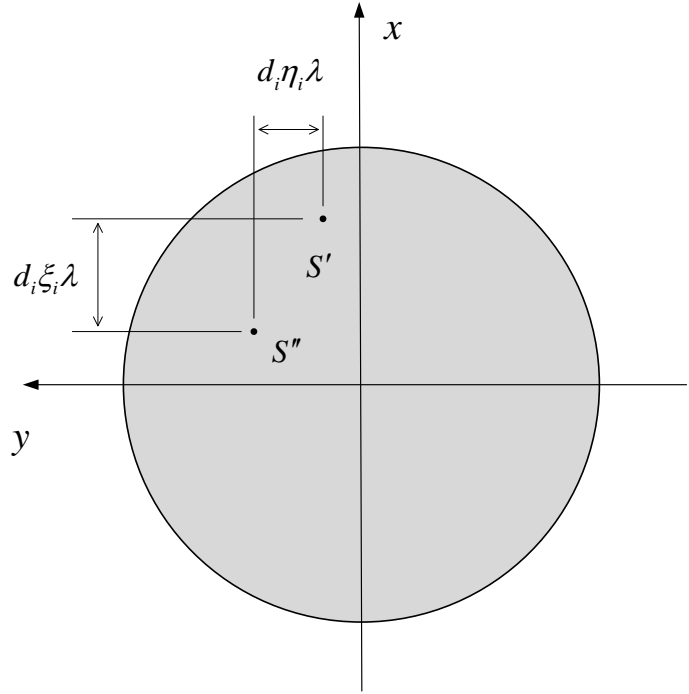

Figure 2. coordinate difference of points $S'$ and $S''$ on xy-plane

In the Fig.1, $x_i = -x_o = -x$, $y_i = -y_o = -y$. So in the $x_i y_i z$ coordinate, the coordinates of $S'$ and $S''$ can be written down as $S'(-x-\lambda d_i \xi_i, -y-\lambda d_i \eta_i)$ and $S''(-x,-y)$. After refracted through the lens, the beams $s'$ and $s''$ propagated along the $S'O_i$ and $S''O_i$ directions respectively. According to the position of $S'$ and $O_i$, the component of $\vec{k}'$ on $x_i$-axis and $y_i$ can be that:

$$k'_{x_i} = \frac{2\pi}{\lambda} \frac{x+\lambda \xi_i d_i}{d_i} \tag{2}$$

$$k'_{y_i} = \frac{2\pi}{\lambda} \frac{y+\lambda \eta_i d_i}{d_i} \tag{3}$$

From the coordinates of $S'$ and $O_i$, the component of $\vec{k}''$ on $x_i$-axis and $y_i$ can be that:

$$k''_{x_i} = \frac{2\pi}{\lambda} \frac{x}{d_i} \tag{4}$$

$$k''_{y_i} = \frac{2\pi}{\lambda} \frac{y}{d_i} \tag{5}$$

According to the expression of $I_o(x_o, y_o)$, the phase of cosine fringe of object plane is 0, which also means that the phase is same when $s'$ and $s''$ leave the $x_o y_o$-plane. When the lens is the ideal lens, such as Fig.1, all the optical distance from the $O_o$ to $O_i$ should be the same.

Then the phase of $s'$ and $s''$ at $O_i$ will be the same, and the amplitude of $s'$ and $s''$ on $x_i y_i$-plane is that:

$$U'(S'; x_i, y_i) = P(S') \exp\left[i(k'_{x_i} x_i + k'_{y_i} y_i)\right] \tag{6}$$

$$U''(S''; x_i, y_i) = P(S'') \exp\left[i(k''_{x_i} x_i + k''_{y_i} y_i)\right] \tag{7}$$

During the equation (6) and equation (7), $P(x, y)$ is the pupil function of the lens[1,9]. The practical lens always has the aberration, if the aberration of lens in Fig.1 is $W(x, y)$, the generalized pupil function of lens will be[10,11.12]:

$$\tilde{P}(x, y) = P(x, y) \exp\left[ikW(x, y)\right] \tag{8}$$

In the equation (8), $k = 2\pi/\lambda$. Using $\tilde{P}(x, y)$ to replace $P(x, y)$ in equation (6) and equation (7), the amplitude of $s'$ and $s''$ on the $x_i y_i$ plane will be that:

$$U'(S'; x_i, y_i) = \tilde{P}(S') \exp\left[i(k'_{x_i} x_i + k'_{y_i} y_i)\right] \tag{9}$$

$$U''(S''; x_i, y_i) = \tilde{P}(S'') \exp\left[i(k''_{x_i} x_i + k''_{y_i} y_i)\right] \tag{10}$$

$s'$ and $s''$ interfered on the $x_i y_i$ plane, and formed cosine fringes:

$$dI_i(x_i, y_i) = |U'(S'; x_i, y_i) + U''(S''; x_i, y_i)|^2 \tag{11}$$

Then we can substitute equation (2) to (5) and (8) to (10) into equation (11), we get that:

$$dI_i(x_i, y_i) = \left\{ |\tilde{P}(S')|^2 + |\tilde{P}(S'')|^2 + \tilde{P}(S')\tilde{P}^*(S'') \exp\left[i2\pi(\xi_i x_i + \eta_i y_i)\right] \right.$$
$$\left. + \tilde{P}^*(S')\tilde{P}(S'') \exp\left[-i2\pi(\xi_i x_i + \eta_i y_i)\right] \right\} dxdy \tag{12}$$

$I_o(x_o, y_o)$ not only emits interference pairs for $s'$ and $s''$, but it also emits an infinite number of interference pairs. During the imaging process of Fig.1, each pair of interference pairs forms a cosine fringe on $x_i y_i$ surface respectively. And the intensity distribution of $x_i y_i$ can be obtained by superimposing each cosine fringe with light intensity. According to the result of equation (12), the phase of $dI_i(x_i, y_i)$ can be determined by the $\tilde{P}(S')$ and $\tilde{P}(S'')$, which means that the phase of fringes can be determined by the $W(S')$ and $W(S'')$.

## 2. The effect of aberration optical transfer function

In the lens imaging process of Figure.1, we superimposed the cosine fringe formed on the surface of $x_i y_i$ by all the interference pairs, so as to obtain the light intensity distribution on the surface of $x_i y_i$. When the lens has no aberration such as Fig.1, the stripes formed by all the interference pairs overlap. The Figure.3 shows that the coincidence of $dI_l$, $dI_m$ and $dI_n$ formed by the interference pairs of *m, n* and *l*. If the lens in Fig.1 has aberration, the stripes formed by each interference pair may not be identical. Fig.4 is the situation where the stripes formed by the three interference pairs do not coincide. After adding the total stripe modulation reduced, the phase also changed.

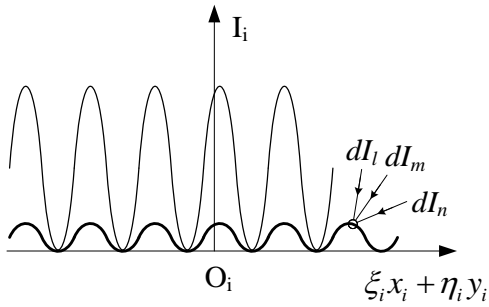 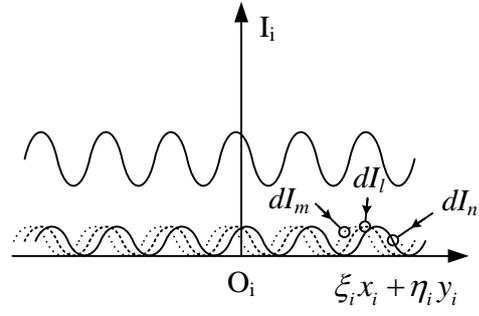

Figure 3. The stripes of an ideal lens    Figure 4. A fringe with an aberration lens

The total stripe brightness on $x_i y_i$ plane can be written down by having the integral of equation (12):

$$I_i(x_i, y_i) = 2S + \exp[i2\pi(\xi_i x_i + \eta_i y_i)] \iint \tilde{P}(S')\tilde{P}^*(S'')dxdy$$

$$+ \exp[-i2\pi(\xi_i x_i + \eta_i y_i)] \iint \tilde{P}^*(S')\tilde{P}(S'')dxdy \quad (13)$$

During the equation (13), $S = \iint |\tilde{P}(S')|^2 dxdy = \iint |\tilde{P}(S')|^2 dxdy$, which is the pupil area. Assuming that:

$$\iint \tilde{P}(S')\tilde{P}^*(S'')dxdy = m(\xi_i, \eta_i)\exp[i\varphi(\xi_i, \eta_i)] \quad (14)$$

$$\iint \tilde{P}^*(S')\tilde{P}(S'')dxdy = m(\xi_i, \eta_i)\exp[-i\varphi(\xi_i, \eta_i)] \quad (15)$$

We can get the brightness of the image by substitute equations (14) and (15) into equation (13), it is that:

$$I_i(x_i, y_i) = 2S + m(\xi_i, \eta_i)\exp[i2\pi(\xi_i x_i + \eta_i y_i)]\exp[i\varphi(\xi_i, \eta_i)]$$

$$+ m(\xi_i, \eta_i)\exp[-i2\pi(\xi_i x_i + \eta_i y_i)]\exp[-i\varphi(\xi_i, \eta_i)]$$

$$= 2S + 2m(\xi_i, \eta_i)\cos[2\pi(\xi_i x_i + \eta_i y_i) + \varphi(\xi_i, \eta_i)]$$

$$= 2S\left\{1 + \frac{1}{S}m(\xi_i, \eta_i)\cos[2\pi(\xi_i x_i + \eta_i y_i) + \varphi(\xi_i, \eta_i)]\right\} \quad (16)$$

From the equation (16), the image plane cosine fringe modulation and the phase can be got which are $\frac{1}{S}m(\xi_i, \eta_i)$ and $e^{i\varphi(\xi_i, \eta_i)}$. Comparing the equation (16) with the equation (1), it can be concluded that lens MTF, PTF and OTF are respectively [8,13]:

$$MTF(\xi_i, \eta_i) = \frac{1}{S}m(\xi_i, \eta_i) \quad (17)$$

$$PTF(\xi_i, \eta_i) = e^{i\varphi(\xi_i, \eta_i)} \quad (18)$$

$$OTF(\xi_i, \eta_i) = MTF(\xi_i, \eta_i)PTF(\xi_i, \eta_i) = \frac{1}{S}m(\xi_i, \eta_i)\exp[i\varphi(\xi_i, \eta_i)] \quad (19)$$

Substituting equation (14) into equation (19) and get:

$$OTF(\xi_i, \eta_i) = \frac{1}{S}\iint \tilde{P}(S')\tilde{P}^*(S'')dxdy$$

$$= \frac{1}{S}\iint \tilde{P}^*(x, y)\tilde{P}(x + \lambda d_i \xi_i, y + \lambda d_i \eta_i)dxdy \quad (20)$$

The results of (20) shows that the OTF of the system is equal to the normalized autocorrelation of the generalized optical pupil function, which is consistent with the classical OTF theory [1] based on mathematical analysis, indicating that the OTF theory without Fourier transform can be applied to optical systems with aberrations. Equation (20) is based on the propagation process of light beam. This method can calculate OTF directly according to the construction of lens. However, for the complex lens construction, the OTF of system can be obtained according to the propagation path of light beam, which is the advantage of the OTF theory without Fourier transform.

## 3. OTF analysis of complex optical systems

Generalized pupil function $\tilde{P}(x, y)$ determines the system OTF. For the single thin lens, the form of generalized pupil function is relatively simple. But for the actual optical system, it is more complex. At this time, the generalized pupil function must be obtained by combining the system construction and the using conditions. In the Fig.5, the optical system consists of two lenses $L_1$ and $L_2$, which has aberrations for each interface. $W_1$ is the aberration of the anterior surface of $L_1$, $W_2$ is the aberration of the posterior surface of $L_1$, $W_3$ is the aberration of the anterior surface of $L_2$, and $W_4$ is the aberration of the posterior surface of $L_2$.

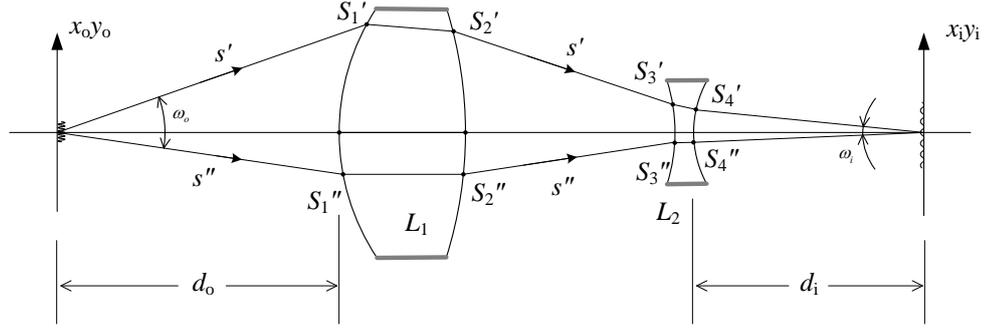

Figure 5. The path of two beams of interference pair passing through the optical system

In the Fig.5 the cosine fringe of the object surface emits interference pairs, $s'$ and $s''$ are two beams of one interference pair. The beam $s'$ goes through $S_1', S_2', S_3'$ and $S_4'$ to $x_i y_i$-plane, and the beam $s''$ through $S_1'', S_2'', S_3''$ and $S_4''$ to $x_i y_i$-plane.

The total aberration of beam $s'$ in the optical system is $W_1(S_1') + W_2(S_2') + W_3(S_3') + W_4(S_4')$, and the total aberration of beam $s''$ in the optical system is $W_1(S_1'') + W_2(S_2'') + W_3(S_3'') + W_4(S_4'')$. The generalized pupil function in Fig.5 is that:

$$\tilde{P}(x_4, y_4) = P(x_4, y_4) \exp[ik(W_1 + W_2 + W_3 + W_4)] \qquad (21)$$

Where the exit pupil coordinates is $(x_4, y_4)$. According to the equation (20), the OTF of optical system in Fig.5 can be:

$$OTF(\xi_i, \eta_i) = \frac{1}{S} \iint \tilde{P}^*(x_4, y_4) \tilde{P}(x_4 + \lambda d_i \xi_i, y_4 + \lambda d_i \eta_i) dx_4 dy_4 \qquad (22)$$

In the actual process, in order to simplify the OTF calculation of the system, a certain number of beams can be selected according to the system construction, the total aberration of each beam can be calculated, the generalized pupil function of the optical system can be fitted out, and then the generalized pupil function can be normalized and autocorrelated, at last the OTF of optical system can be obtained. When calculating the generalized pupil function, the more the number of beams is selected, the more detailed the generalized pupil function is obtained, and the more accurate the OTF is calculated. The imaging process of the cosine fringe of the proximal axis is introduced. For the cosine fringe beyond the optical axis, the generalized pupil function and OTF can also be calculated according to the above method.

## Conclusion

The essence of non-Fourier transform OTF theory is the imaging theory of frequency domain, which has the intuitiveness of geometrical optics and the accuracy of wave optics. The non-Fourier transform OTF theory is based on the interference pair. Under the condition of incoherent imaging, all interference pairs have equally probability. According to the propagation path of the interference pair in the imaging process, the OTF of the lens is obtained. And we found that the pupil of the lens affects the utilization efficiency of the interference pair and the aberration of the lens affects the quality of the interference pair. The non-Fourier transform OTF theory replaces the geometrical optics with wave. It can analyze the complex optical system, providing help for the optical system design and imaging quality evaluation.

## Acknowledgement

We thank Qi Lu for her advice and assistance with the paper writing.